\documentclass[entropy,article,submit,pdftex,moreauthors]{Definitions/mdpi} 
\firstpage{1} 
\pubvolume{1}
\issuenum{1}
\articlenumber{0}
\pubyear{2025}
\copyrightyear{2025}
\datereceived{ } 
\daterevised{ } % Comment out if no revised date
\dateaccepted{ } 
\datepublished{ } 
\hreflink{https://doi.org/} % If needed use \linebreak
\usepackage{bm}
\usepackage{bbm}
\usepackage{float}
\usepackage{graphicx,wrapfig}% Include figure files
\usepackage{soul}
\usepackage{verbatim}
\usepackage{bbm}
\usepackage{braket}
\usepackage{enumitem}
\usepackage{qcircuit}
\usepackage[normalem]{ulem}
\usepackage{xcolor}

\def\ep{\varepsilon}

\def\rb{{\bf r}}

\def\ep{\varepsilon}
\def\qb{{\bf q}}

\newcommand{\tr}{\mathrm{Tr}}

\def\0{^{(0)}}
\def\1{^{(1)}}

\Title{Electrons in quantum dots on helium: from charge qubits to synthetic color centers}

\TitleCitation{Title}

\Author{M. I. Dykman $^{1}$\orcidA{0000-0003-3996-7932} and J. Pollanen $^{1}$ \orcidB{0000-0002-4314-5045}}

\AuthorNames{Firstname Lastname, Firstname Lastname and Firstname Lastname}

\isAPAStyle{%
       \AuthorCitation{Dykman, M. I. \& Pollanen, J.}
         }{%
        \isChicagoStyle{%
        \AuthorCitation{Dykman, M. I. \& Pollanen, J.}
        }{
        \AuthorCitation{Dykman, M. I. \& Pollanen, J.}
        }
}

\address[1]{%
$^{1}$ \quad Department of Physics and Astronomy, Michigan State University, East Lansing, MI 48824, USA; dykmanm@msu.edu, pollanen@msu.edu}
\abstract{Electrons trapped above the surface of helium provide a means to study many-body physics free from the randomness that comes from defects in other condensed-matter systems. Localizing an electron in an electrostatic quantum dot makes its energy spectrum discrete, with controlled level spacing. The lowest two states can act as charge qubit states. In this paper, we study how the coupling to the quantum field of capillary waves on helium — ripplons — affects electron dynamics. As we show, the coupling can be strong. This bounds the parameter range where  electron-based charge qubits can be implemented. The constraint is different from the conventional relaxation time constraint. The electron-ripplon system in a dot is similar to a color center formed by an electron defect  coupled to phonons in a solid. In contrast to solids, the coupling in the electron on helium system can be varied from strong to weak. This enables a qualitatively new approach to studying color center physics. We analyze the spectroscopy of the pertinent synthetic color centers in a broad range of the coupling strength.}


\begin{document}

%%%%%%%%%%%%%%%%%%%%%%%%%%%%%%%%%%%%%%%%%%

This paper is part of the collection in honor of  P.~V.~E.~McClintock. Peter~McClintock is not only an outstanding scientist, with fundamental contributions to many areas of science, but also a remarkable human being. One of us (M.D.) has the privilege of counting Peter as a personal friend and has come to admire not only the depth of Peter's thinking, but also his kindness and warmth, his attention to not only scientific but also personal matters, and the courage to stand up for what is right. 

\section{Introduction}
\label{sec:intro}

Studying the two-dimensional (2D) system of electrons on the surface of liquid helium is interesting in several aspects. In this system electrons are strongly interacting with each other and are coupled to a quantum field of helium vibrations, which leads to rich and nontrivial many-body behavior. At the same time, the system is pristine, with no defects, allowing one to investigate this behavior in a controlled way, which is hard if not impossible to do in other areas of condensed matter physics. In particular, the electron-electron interaction can be controlled by varying the electron density by several orders of magnitude, from $10^{11}~\mathrm{cm}^{-2}$ down to $10^6~\mathrm{cm}^{-2}$. The strength of the coupling to the helium vibrations can be also controlled over a broad range. This coupling is determined by the electric field $E_\perp$, which  presses the electrons against the helium surface and can be changed in the experiment from $\sim 10^2$~V/cm to $\sim 10^4$~V/cm. 

In the overwhelming majority of experiments conducted so far the electron system on helium was nondegenerate, the electron wave functions did not overlap. Yet the system is strongly correlated. The correlations are at the root of many phenomena observed in the system. These phenomena range from  Wigner crystallization, which was first observed with electrons on helium~\citep{Grimes1979,Fisher1979}, to many-electron tunneling~\citep{Menna1993,Dykman2001b}, magnetoconductivity and cyclotron resonance~\citep{Dykman1979a,Edelman1979,Iye1980,Wilen1988a,Dykman1993b,Lea1994a,Monarkha1997,Mistura1998,Klier2002}, radio-frequency, microwave and piezoacoustic response~\citep{ANDREI1994,Konstantinov2009,Konstantinov2010,Konstantinov2013,Yunusova2019,Chepelianskii2021,Kawakami2021,Byeon2021,Li2024a,He2025,Jennings2025,Mikolas2025}, and profound nonlinear effects~\citep{Kristensen1996,Dykman1997b,Vinen1999,Glasson2001, Collin2002,Ikegami2009,Rees2012,Chepelianskii2015,Rees2016,rees2016a,Siddiq2023}. Interestingly, but not unexpectedly, the electron-electron interaction is competing with the coupling to the vibrational excitations in the helium. The latter coupling can lead to a polaronic effect~\citep{Shikin1974a}, which can be strong, particularly in a magnetic field~\citep{Dykman1978a}. However, because of thermal density fluctuations in  the electron liquid, an electron can be ``blown away'' from the polaronic  well~\citep{Dykman1979a}. In particular, it is this effect that is behind the observed magnetoconductivity, which is very different from the conventional magnetoconductivity of other 2D electron systems~\citep{Dykman1993b}.

In spite of the significant progress in understanding many aspects of the physics of electrons on helium, the fundamental question of the single-electron polaronic effect and its consequences remains unanswered. In this paper we study this effect in the setting where the electron states can be well-controlled. Moreover, we show that, with electrons on helium, it becomes possible to implement, using the polaronic coupling, a ``tunable synthetic  color center''. The importance of such an implementation follows from the fact that color centers are a prominent type of defects in solids~\cite{Stoneham2001}. These defects  have been attracting increasing attention recently in the context of quantum measurements and quantum information~\cite{Maze2008,Grinolds2013,Myers2014,Becker2018,Awschalom2018,Chen2019,Barry2020,Bhaskar2020,Huxter2022,Xu2023a,Harris2024}. However, in solids, the coupling between electronic transitions in color centers and phonons cannot be tuned, which limits the analysis. In contrast, this limitation does not exist for electrons on helium.

\begin{figure}[H]
%\isPreprints{\centering}{} % Only used for preprints
%\begin{center}
\centering{\includegraphics[width=6 cm]{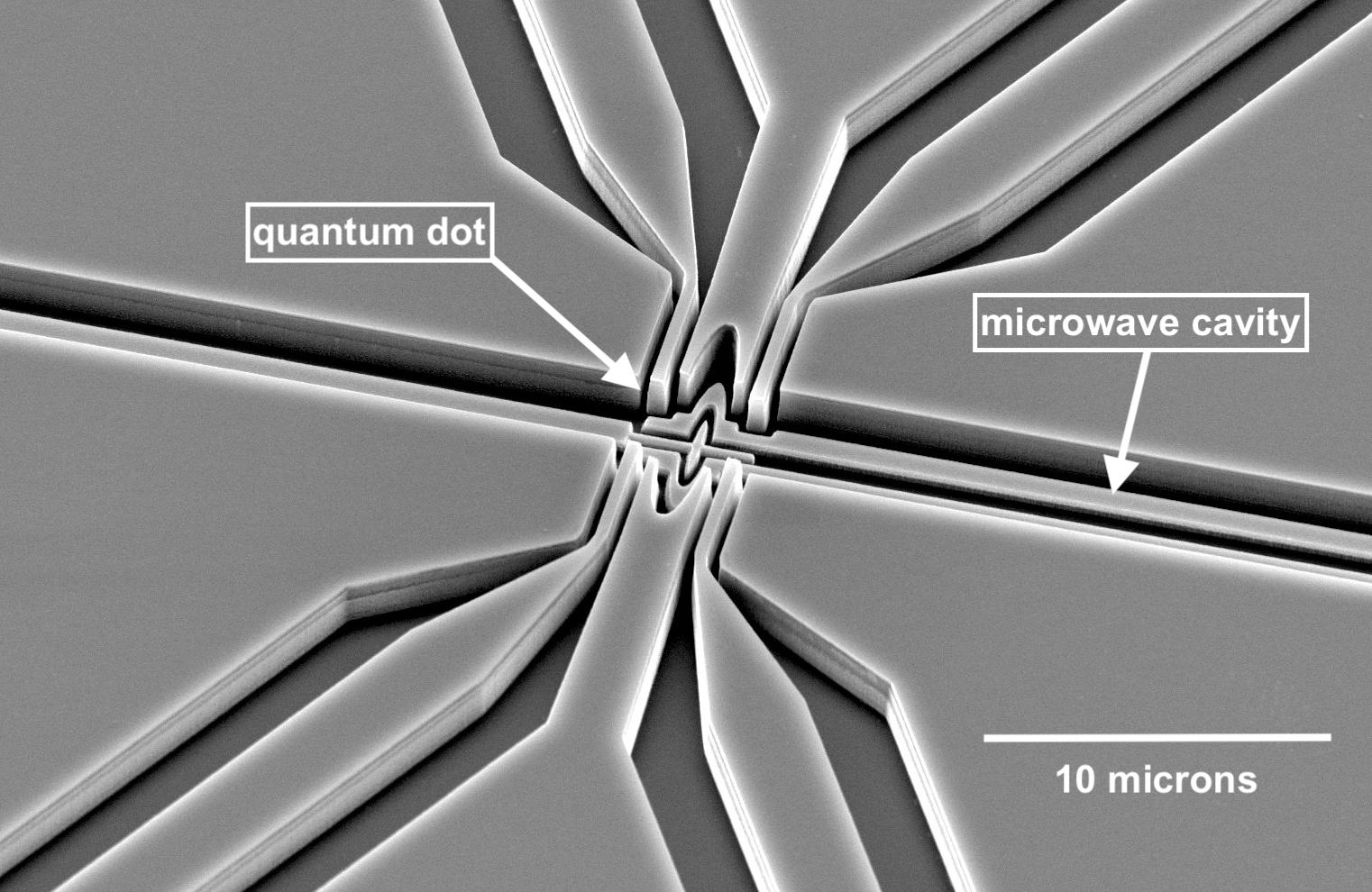}}
%\end{center}
\caption{Electron in helium quantum dot device similar to the devices described in Ref.~\cite{Koolstra2019a}. Trenches of order $0.5-1~\mu$m are dry-etched into a high-resistivity silicon to confine superfluid helium via capillary action. Electrostatic gates are defined by patterning of a pre-sputtered niobium film. They enable tuning of $E_\perp$ which, in turn, enables a controlled polaronic coupling between the electron and the bosonic excitations of helium. The geometrically and electrostatically defined dot is coupled to a microwave resonator cavity to perform spectroscopy measurements of the states of the trapped electron.\label{fig1}}
\end{figure}

A theoretical study of the single-electron polaronic effect has been made timely by the recent progress in the experimental techniques. It is now possible to place the electron system into a high-quality-factor microwave cavity~\cite{Yang2016a,Koolstra2025,Belianchikov2025}. Quantum measurements of the cavity response enable accessing features of the electron dynamics at the single-electron level, that have not been possible in the past. In parallel, precision methods have been developed for single-electron confinement in electrostatically created quantum dots on the helium surface, which can be embedded into micro-cavities~\citep{Koolstra2019a,Castoria2024}. A representative example of such a dot placed into a microwave cavity is shown in Fig.~\ref{fig1}.

In this paper we study the response of an electron confined in a quantum dot to a resonant field that causes transitions between quantized intradot energy levels. These transitions are analogous to the transitions between the electron states of an electron localized at a defect in solid, i.e., between the states of a color center. We show that, by varying the strength of the coupling of the localized electron on helium to the helium vibrational modes, one can reproduce and explore various regimes encountered in the physics of color centers, from strong to intermediate to weak coupling. 

Another motivation for this study comes from the idea of using electrons in quantum dots on helium as charge qubits~\cite{Platzman1999,Dykman2003a,Schuster2010,Pollanen2021,beysengulov2023}. Critical for implementing a quantum computer based on electrons on helium is understanding of the polaronic effect. In the standard terms of quantum information~\citep{Nielsen2011} this effect is an analog of dephasing. However, it is far more complicated and is not described by a dephasing rate. We will study how the polaronic effect is manifested in the gate operations on a charge qubit based on an electron in a quantum dot on helium. Various aspects of this effect, including the ultimate limit on the gate fidelity will be analyzed.

%%%%%%%%%%%%%%%%%%%%%%%%%%%
%%%%%%%%%%%%%%%%%%%%%%%%%%%%%%%%%%%%%%%%%%

\section{Model}
\label{sec:model}

We consider a geometry where electrons are floating in the $x-y$ plane above the helium surface. They are bound to the surface by an image potential $-\Lambda/z$ for $z>0$, where $\Lambda= (\ep -1)e^2/4(\ep+1)$ with $\ep \simeq 1.057$ being the helium dielectric constant; the coordinate $z$ is counted  off the surface. There is a $\sim 1$~eV exchange-force barrier preventing  penetration into the helium. The motion of the electron normal to the surface is therefore quantized, and the spectrum is Rydberg-like, with the characteristic Bohr radius $r_B = \hbar^2/\Lambda m_e \approx 76$\AA. The spacing between the lowest and the first excited states is $\gtrsim 120$~GHz in frequency units; it largely exceeds the relevant temperatures and the energies of in-plane motion~\citep{Shikin1974a,Cole1974,Grimes1976a}. 

Quantum dots are created by electrodes submerged in the helium at a  depth $d_\mathrm{dot}\simeq 0.5~\mu$m~\cite{Dykman2003a,Schuster2010,Koolstra2019a}. The low-lying states of intradot electron motion are weakly non-equidistant states of two vibrational modes. We will choose the coordinates of these modes to be pointing along the $x$ and $y$ axes and call them $x$ and $y$ modes, respectively. The electron states then are the Fock states of two oscillators $\ket{n_x,n_y}$, where $n_{x,y}=0,1,2,...$. The mode eigenfrequencies  $\Omega_x$ and $\Omega_y$ 
are controlled by the electrode potential. The Hamiltonian that describes the low-energy in-plane intradot states reads
\begin{align}
\label{eq:H0}
&H_0=\sum_{i=x,y}\hbar\Omega_i a_i^\dagger a_i + H_\mathrm{nonlin}\,\qquad H_\mathrm{nonlin} = \frac{1}{2}\sum_{i=x,y}\hbar V_{ii} a_i^\dagger{}^2 a_i^2 + \hbar V_{xy}a_x^\dagger a_x a_y^\dagger a_y.
\end{align}
Here $a_i$, $a_i^\dagger$ ($i=x,y$) are the ladder operators of the $x$ and $y$ modes. The parameters $V_{ii}$ and $V_{xy}$ describe the internal mode nonlinearities and the cross-nonlinearity, respectively. For typical quantum dots they are small compared to $\Omega_{x,y}$. They determine the non-equidistance of the energy levels of the intradot vibrations. The smallness of the nonlinearity parameters is a consequence of the smallness of the quantum localization lengths $l_i=(\hbar/2m_e\Omega_i)^{1/2}$ ($i=x,y$) compared to the size of the dot; here $m_e$ is the electron mass.   

Making the dot potential asymmetric in the $x$ and $y$ directions allows one to make the frequencies $\Omega_x$ and $\Omega_y$ different, so that $|\Omega_x - \Omega_y| \gg |V_{xx}|, |V_{yy}|, |V_{xy}|$. Typically, for electron dots in superconducting microwave cavities, of interest are frequencies $\Omega_{x,y}$ in the range of several gigahertz. We will assume that the temperature is low, $k_BT \ll \hbar\Omega_{x,y}$, so that the electron is in the ground intradot state. Given the above estimate of the frequencies, this corresponds to temperatures $\lesssim 0.1-0.5$~K, which are routinely used in experiments on electrons on helium.

The relevant excitations in  helium are the capillary waves, ripplons, and acoustic phonons. The phonons play an important role in the electron energy relaxation. In terms of the polaronic effects, of primary interest are ripplons, which typically have low frequencies and are more strongly coupled to surface electrons~\citep{Shikin1974a,Cole1974}. These surface waves are characterized by their 2D wave vector $\qb$. The Hamiltonian $H_\mathrm{int}$ of the electron-ripplon coupling and the ripplon Hamiltonian $H_r$ have the form
\begin{align}
\label{eq:Hi}
 H_\mathrm{int}=\sum_\qb \left(V_\qb e^{i\qb\rb} b_\qb + V_\qb^* e^{-i\qb\rb} b^\dagger_\qb\right),\qquad
H_r=\hbar\sum_{\qb,}\omega_q b^\dagger_\qb b_\qb,
\end{align}
where $\rb = (x,y)$ is the 2D electron coordinate, $V_\qb$ are the coupling parameters, $\omega_q$ is the frequency and $b_\qb$ is the annihilation operator of the ripplon with wave vector $\qb$. Since helium is isotropic, $\omega_q$ is independent of the direction of $\qb$. 

The values of $V_\qb$ are obtained by projecting the overall coupling energy onto the lowest state of electron motion normal to the surface. They have the form~\cite{Shikin1974a}
\begin{align}
    \label{eq:V_q}
 &V_\qb = (V_\qb^\mathrm{(pol)} + eE_\perp )(\hbar q/2\rho_\mathrm{He}\omega_q S_\mathrm{He})^{1/2}, \quad 
    V_\qb^\mathrm{(pol)}%\approx -\frac{\Lambda q^2}{2}[\ln(qr_B/4)+1],\nonumber\\
    \approx% -2\Lambda q^2 \nonumber\\ 
    %&\times \left[Q_B -2\sqrt{Q_B}\ln[(\sqrt{Q_B}+2)/qr_B]\right]/Q_B^2, \quad Q_B=4-(qr_B)^2.
    -(\Lambda q^2/2)\left[1+ \ln (qr_B/4)\right].
\end{align}
Here, $V_\qb^\mathrm{(pol)}$ comes from  the modulation of the image potential by surface waves. The term $\propto eE_\perp$ comes from the change of the electron energy in the pressing field $E_\perp$ due to rising and lowering of the helium surface, and thus the electron, by surface waves; $\rho_\mathrm{He}$ is the helium density and $S_\mathrm{He}$ is the helium surface area. We use the expression for $V_\qb^\mathrm{(pol)}$ that applies for the variational wave function of motion normal to the surface of the form $\psi(z)\propto z \exp(-z/r_B)$. Of interest is the case where the Bohr radius $r_B$ is small compared to typical value of $q^{-1}$, and we have kept the leading-order terms in $qr_B$; $r_B$ becomes a variational parameter for $E_\perp >0$, cf.~\cite{Dykman2003a}.

%%%%%%%%%%%%%%%%%%%%%%%

\subsection{The adiabatic approximation}
\label{subsec:adiabatic}

The wave numbers $q$ of the ripplons coupled to an intradot electron are effectively limited by the reciprocal electron localization lengths $l_x^{-1},l_y^{-1}$. This is seen from Eq.~(\ref{eq:adiabatic}) below and can be easily understood, since the effect of ripplons with wavelengths small compared to $l_x,l_y$ is averaged out. The typical frequencies of the relevant ripplons are 
\begin{align}
    \label{eq:typical_omega_q}
    \omega_q\lesssim \omega_{q_m}~, \qquad q_m = (l_x^2 + l_y^2)^{-1/2}.
\end{align}
They are much smaller than the electron vibration frequencies $\Omega_{x,y}$. For typical $\Omega_{x,y}/2\pi$ in the range of $3 - 6$~GHz, we have $\omega_{q_m} \simeq 1.2\times 10^8 - 2.1\times 10^8$~s$^{-1}$. As a consequence, there is no single-ripplon decay of the vibrational electron states. The main effect of the coupling to ripplons is the ripplon-induced modulation of the electron energy levels. Such modulation can be described in the adiabatic approximation. In this approximation one keeps in the electron operator $\exp(i\qb\rb)$ in $H_\mathrm{int}$, Eq.~(\ref{eq:Hi}), only the terms that are diagonal in the intradot states, whereas the off-diagonal terms are disregarded. Respectively, the coupling Hamiltonian takes the form $H_\mathrm{int}^\mathrm{(ad)}$, where 
\begin{align}
    \label{eq:adiabatic}
    H_\mathrm{int}^\mathrm{(ad)} = \sum_\qb 
    %e^{i\qb\rb}\Rightarrow %\exp\left(-\frac{1}{2}\sum_{i=x,y}l_i^2q_i^2\right)
    e^{-\eta_\qb}\sum_{n,m}\frac{(il_xq_x)^{2n}}{n!^2}a_x^\dagger{}^n a_x^n \, \frac{(il_yq_y)^{2m}}{m!^2}a_y^\dagger{}^m a_y^m (V_\qb b_\qb + V_\qb^* b_\qb ^\dagger), \qquad \eta_\qb = \frac{1}{2}\sum_{i=x,y}l_i^2q_i^2.
\end{align}
Here we used the standard expression for  the components $x,y$ of the electron displacement operator, $x = l_x(a_x^\dagger + a_x)$ and  $y=l_y(a_y^\dagger + a_y)$.  The nonadiabatic terms, which are disregarded in the replacement $H_\mathrm{int} \to H_\mathrm{int}^\mathrm{(ad)}$, lead to small corrections to the frequencies $\Omega_x,\Omega_y$.

The total Hamiltonian of the electron-ripplon system in a dot on helium in the adiabatic approximation is
\begin{align}
    \label{eq:full_H}
H=H_0 + H_\mathrm{int}^\mathrm{(ad)} + H_r.
\end{align}

%It is seen from Eq.~(\ref{eq:adiabatic}) that the intradot electron is coupled to ripplons with wave numbers $q\lesssim l_x^{-1}, l_y^{-1}$. Frequencies of such ripplons are typically small. For $\Omega_{x,y}/2\pi$ in the range of $3 - 6$~GHz, we have $\omega_q \sim 1.2\times 10^8 - 2.1\times 10^8$~s$^{-1}$.   
The smallness of the characteristic ripplon frequencies $\omega_q$ compared to the frequencies  $\Omega_{x,y}$ of the intradot electron excitations points to a  similarity with the physics of color centers. The electron transition frequencies of color centers are often in the visible or near-infrared range~\cite{Stoneham2001,Awschalom2018,Barry2020,Bhaskar2020},  largely exceeding the phonon frequencies. The coupling to phonons in color centers is determined by the structure of the underlying defects and is often strong. However, it is not tunable \emph{in situ}, whereas the coupling of an intradot electron to ripplons can be controlled by varying $E_\perp$, as seen from Eqs.~(\ref{eq:Hi}) and (\ref{eq:V_q}). We  note that the shape of the spectra of color centers is also usually analyzed in the adiabatic approximation, cf.~\cite{Pekar1950,Huang1950,Stoneham2001}.

%%%%%%%%%%%%%%%%%%%%%%%%%%%%
%%%%%%%%%%%%%%%%%%%%%%%%%%%%%

\section{Resonant linear response of intradot electrons}
\label{sec:transitions}

Electron transitions between the intradot vibrational states have a large dipole moment $|e\rb| \sim e l_{x,y}$. Therefore they can be comparatively strongly coupled to the intracavity microwave field (this, in fact, is what enables detecting single electrons using a microwave cavity). For temperatures $k_BT\ll \hbar\Omega_{x,y}$ the electron occupies the ground vibrational state $\ket{n_x,n_y}$ with $n_x=n_y=0$. By tuning one of the transition frequencies, for example $\Omega_x$, close to the cavity eigenfrequency one can study the spectrum of resonant response to a microwave field associated with the transition $\ket{0_x,0_y}\to \ket{1_x,0_y}$. 

The response to a weak microwave field at frequency $\omega$ is characterized by the electron conductivity. 
Using the Kubo formula, the diagonal component of the electron conductivity for a cavity field polarized along the $x$-axis can be written as 
\begin{align}
\label{eq:Kubo}
\sigma_{xx}(\omega) =\frac{e^2\omega}{\hbar}\,
\int_0^\infty dt e^{i\omega t}\langle [x(t),x(0)]\rangle\,,
\end{align}
where $\braket{\cdot}$ implies thermal averaging.

The real part of $\sigma_{xx}$ gives the absorption coefficient of the field. For $\omega\approx\Omega_x$ and in the case where the linewidth of the resonant peak of Re~$\sigma_{xx}$ is small compared to $\Omega_x$ and $ |\Omega_x-\Omega_y|$, we can keep only the term $\propto \exp(-i\Omega_x t)$ in the expression for $\braket{[x(t),x(0)]}$. For low temperatures and $\omega\approx \Omega_x$ this gives
\begin{align}
\label{eq:absorption_general}
\mathrm{Re}\,\sigma_{xx}(\omega) \approx \frac{e^2}{2m_e}\alpha_{xx}(\omega)\,, \qquad \alpha_{xx}(\omega)=\mathrm{Re}\,
\int_0^\infty dt e^{i\omega t}\braket{a_x(t)a_x^\dagger(0)}.
\end{align}
If there were no coupling to ripplons, we would have $\alpha_{xx}(\omega) = \pi \delta(\omega-\Omega_x)$, i.e., the cavity absorption spectrum would have the form of a $\delta$-peak at the intradot frequency $\Omega_x$. The electron-ripplon coupling leads to broadening of this spectroscopic peak. We note that the mode nonlinearity and the nonlinear mode coupling do not affect the spectral peak for low temperatures.

%%%%%%%%%%%%%%%%%%%%%%%%%%%%%%%%%

\subsection{Averaging over the ripplon states}
\label{subsec:averaging_ground}

The averaging in Eq.~(\ref{eq:absorption_general}) implies a trace over the electron and ripplon states with the weight $Z^{-1}\exp(-\beta H)$, where $H$ is the full Hamiltonian of the system, see Eq.~(\ref{eq:full_H}), $\beta=1/k_BT$, and $Z$ is the partition function. For typical temperatures $T~>~10$~mK used in the experiments, the averaging should be done assuming that the ripplons coupled to the electron are thermally excited even though the  electron itself is in the ground intradot state. 

The electron-ripplon coupling leads to a displacement of the ripplon equilibrium positions, i.e., deformation of the helium surface. If the electron is in the ground state, one can allow for this displacement by making a unitary transformation 
\begin{align}
\label{eq:canonical}
U_g=\exp\left[\sum_\qb \frac{\exp(-\eta_\qb)}{\hbar\omega_q}\, \,(V_\qb b_\qb - V_\qb^* b_\qb^\dagger)\right], 
\end{align}
where we have taken into account the explicit form of the diagonal matrix element of the coupling Hamiltonian $H_\mathrm{int}^\mathrm{(ad)}$,  Eq.~(\ref{eq:adiabatic}), on the electron wave function $\ket{m_x,n_x}$ with $m_x=n_x=0$. 

The transformation (\ref{eq:canonical}) leads to a shift $b_\qb \to b_\qb - V_\qb^* \exp(-\eta_\qb)/\hbar\omega_q$. When substituted into $H_\mathrm{int}^\mathrm{(ad)}$, this shift results in an adiabatic polaronic change of the frequencies and nonlinearity parameters of the electron vibrational modes. In particular, the frequency of the $x$-mode is shifted from $\Omega_x$ to $\Omega_\mathrm{ad}$,
\begin{align}
    \label{eq:x_mode_renomralize}
    \Omega_x \to \Omega_\mathrm{ad}, \quad \Omega_\mathrm{ad} =\Omega_x +P_\mathrm{ad},\qquad P_\mathrm{ad}= 2\sum_\qb (l_x q_x)^2 |V_\qb|^2e^{-2\eta_\qb}/\hbar^2\omega_q.
\end{align}
The shift is quadratic in the coupling parameters and is independent of temperature.

It is convenient to calculate the correlator $\braket{a_x(t) a_x^\dagger (0)}$ in Eq.~(\ref{eq:absorption_general}) by changing to the interaction representation with the Hamiltonian $\tilde H_0+H_r$, where $\tilde H_0$ differs from $H_0$ in that $\Omega_x$ is replaced by $\Omega_\mathrm{ad}$. Then the time evolution operator becomes $U_g^\dagger\exp[-iHt]U_g =\exp[-i(\tilde H_0 +H_r)t]\mathcal{T}_t\exp[-i\int_0^t dt_1 \tilde H_\mathrm{int}^\mathrm{(ad)}(t_1) dt_1]$, where $\mathcal{T}_t$ is the time ordering operator and $\tilde H_\mathrm{int}^\mathrm{(ad)}$ is given by Eq.~(\ref{eq:adiabatic}) in which the sum over $m,n$ runs over $m+n>0$ (the term with $m=n=0$ in $H_\mathrm{int}^\mathrm{(ad)}$ has been eliminated by the canonical transformation). Tracing out the ripplonic variables in a standard way, we obtain 
%Then, writing $\exp[-i\int dt_1 \tilde H_\mathrm{int}^\mathrm{(ad)}(t_1)]$ as a product over the ripplons, from Eq.~(\ref{eq:adiabatic}) we obtain
%
\begin{align}
    \label{eq:correlator_explicit}
    &\braket{a_x(t) a_x^\dagger (0)} %e^{-i\Omega_\mathrm{ad} t}\mathcal{T}_t\prod_\qb (\bar n_q + 1)^{-1} \mathrm{Tr}_\qb\exp\left[i(l_x q_x)^2e^{-\eta_\qb}\int_0^t dt_1\left(V_\qb b_\qb(t_1) + \mathrm{H.c.} 
%    \right)/\hbar\right]\nonumber\\
 %   &\qquad \times\exp(-\beta\hbar\omega_q b_\qb^\dagger b_\qb)/(\bar n_q + 1) %\nonumber\\
   % &
   = e^{-i\Omega_\mathrm{ad} t} \exp[-W(t)],\nonumber\\
    &W(t) = \sum_\qb \frac{|V_\qb|^2}{(\hbar\omega_q)^2} e^{-2\eta_\qb} \,(l_x q_x)^4 \left[(\bar n_q +1)\left(1-e^{-i\omega_q t}\right) + \bar n_q \left(1-e^{i\omega_q t}\right)-i\omega_q t\right]
\end{align}
Here, %Tr$_\qb$ means trace over the states of the ripplonic mode with the wave vector $\qb$; 
$\bar n_q$ is the thermal occupation number of a ripplon with the wave number $q$, 
\[\bar n_q \equiv \bar n(\omega_q), \qquad \bar n(\omega) = [\exp(\hbar\omega/k_B T)-1]^{-1}.\]
%
%In Eq.~(\ref{eq:correlator_explicit}) we used that, in the interaction representation, $b_\qb(t) = b_\qb\exp(-i\omega_q t)$.

%%%%%%%%%%%%%%%%%%%%%%%%%%%%%%%%%%%%%%

\subsection{Radiation emission from the excited state}
\label{subsec:luminescence}

If the intradot electron is excited, it can emit radiation by going from the excited to the ground state. We will consider the radiation spectrum assuming  that it is the $x$-mode that is excited and that the ripplons are in thermal equilibrium. This means that they have ``adjusted'' to the excited state of the mode, while the mode itself is not in thermal equilibrium; for example, it has absorbed an $x$-polarized photon, which has brought it into the first excited vibrational state, where it stays longer than it takes for the ripplons to thermalize.  

The general expression for the electron emission spectrum can be found in a standard way by studying the linear response of the intradot electron to a quantized $x$-polarized intracavity radiation field. Near its maximum, the shape of the spectrum is given by the function 
\begin{align}
    \label{eq:emission_spectral}
\tilde\alpha_{xx}(\omega)= \mathrm{Re}\,\int_0^\infty dt e^{i\omega t}\, \overline{a_x^\dagger(0) a_x(t)}, 
\end{align}
where we use an overline to indicate the averaging described above. 

For emission from the first excited vibrational state of the $x$-mode, a calculation similar to the one in the analysis of the absorption spectrum  gives
\begin{align}
    \label{eq:correlator_emision}
  \overline{a_x^\dagger(0) a_x(t)} =  e^{-i\Omega_\mathrm{ad} t} \exp\left[-W^*(t)+ 2it\sum_\qb  \frac{|V_\qb|^2}{\hbar^2\omega_q} e^{-2\eta_\qb} \,(l_x q_x)^4\right] 
\end{align}
Measuring emission into the cavity thus provides a direct way to detect that the intradot electron was excited. Such detection is demanding, as it requires single-photon resolution. In what follows we focus on the absorption and control of electron transitions by short resonant pulses supplied to the cavity.

%%%%%%%%%%%%%%%%%%%%%%%%%
%%%%%%%%%%%%%%%%%%%%%

\section{Absorption spectrum in the limiting cases}
\label{sec:spectrum_general}

Equations (\ref{eq:absorption_general}) and (\ref{eq:correlator_explicit}) provide an explicit general expression for the absorption spectrum of microwave radiation by an electron in a quantum dot on helium. The expression is simplified in the limiting cases of strong and weak coupling to the ripplons. The strong-coupling condition is
\begin{align}
    \label{eq:strong_coupling}
    \gamma\gg \omega_{q_m}~, \qquad \gamma= \hbar^{-1}\left[\sum_\qb |V_\qb|^2 e^{-2\eta_\qb} \,(l_x q_x)^4 (2\bar n_q +1)\right]^{1/2}.
\end{align}
Physically, this condition means that the coupling energy, $\hbar\gamma$, is much stronger than the typical ripplonic energy $\hbar\omega_{q_m}$. When this condition holds, one can expand $W(t)$ in Eq.~(\ref{eq:correlator_explicit}) to second order in $\omega_q t$, which gives $W(t) \approx \gamma^2 t^2/2$. Then, from Eqs.~(\ref{eq:absorption_general}) and (\ref{eq:correlator_explicit}), the spectrum $\alpha_{xx}(\omega)$ has the shape of a Gaussian peak centered at $\Omega_\mathrm{ad}$,
\begin{align}
    \label{eq:Gaussian_peak}
    \alpha_{xx}(\omega) \approx (\pi/2\gamma)^{1/2}\exp\left[-(\omega - \Omega_\mathrm{ad})^2/2\gamma^2\right],\quad \gamma\gg \omega_{q_m}.
\end{align}
The Gaussian shape of the absorption peak is familiar from the theory of strongly coupled color centers~\cite{Pekar1950,Huang1950,Markham1959}. The characteristic width of the peak $\gamma$ is linear in the coupling strength. 

For not too strong coupling, a very narrow zero-ripplon line emerges on the background of the broad Gaussian peak. This spectral feature is an analog of the zero-phonon lines in the spectra of color centers~\cite{Krivoglaz1953,Stoneham2001} and also an analog of the very narrow lines in M\"ossbauer spectra. The form of this line, $\alpha_\mathrm{zr}(\omega)$, is determined by the behavior of the function $W(t)$ for large $\omega_{q_m}t$. In particular, the position of the line is determined by the last term in Eq.~(\ref{eq:correlator_explicit}) for $W$. In the approximation where we disregard processes leading to transitions between the intradot vibrational states, the zero-ripplon line has the form of a $\delta$-function,
\begin{align}
    \label{eq:zero_ripplon}
    &\alpha_\mathrm{zr}(\omega) = \pi \exp(-\bar W) \delta(\omega-\Omega_\mathrm{zr}), \qquad \Omega_\mathrm{zr} = \Omega_x + P_\mathrm{zr}, \nonumber\\ 
        & \bar W = \sum_\qb \frac{|V_\qb|^2}{(\hbar\omega_q)^2} e^{-2\eta_\qb} \,(l_x q_x)^4 (2\bar n_q +1),\quad P_\mathrm{zr} = \sum_\qb \frac{|V_\qb|^2}{\hbar^2\omega_q} e^{-2\eta_\qb} \,(l_x q_x)^2 [2-(l_xq_x)^2].
\end{align}

The zero-ripplon shift $P_\mathrm{zr}$ of the spectral line (\ref{eq:zero_ripplon})   away from the ``bare'' frequency $\Omega_x$ is independent of temperature. Importantly, it differs from the adiabatic line shift $P_\mathrm{ad}$. This means that the zero-ripplon line is shifted away from the position of the Gaussian peak (\ref{eq:Gaussian_peak}). The factor $\exp(-\bar W)$ is the analog of the Debye-Waller factor in the theory of  x-ray and neutron scattering in solids. To the order of magnitude, $\bar W \sim \gamma^2/\omega^2_{q_m}$. Therefore the zero-ripplon line  (\ref{eq:zero_ripplon}) has an exponentially small intensity (i.e, the spectral peak has an exponentially small area) in the limit of very strong coupling, but if the parameter $\bar W$ is not too large, the line should be clearly resolved on the background of the Gaussian peak.  

In the opposite limit, i.e. for weak electron-ripplon coupling, where $\bar W\ll 1$, the zero-ripplon line is the most intense line in the spectrum. The coupling to ripplons, besides  the shift of the line from $\Omega_x$, leads  to the onset of sidebands, i.e., broad absorption bands on the higher- and lower-frequency sides of the zero-ripplon line. This is again similar to the optical spectra of impurities in solids~\cite{Krivoglaz1953,Stoneham2001}. From Eqs.~(\ref{eq:absorption_general}) and (\ref{eq:correlator_explicit}), to first order in $\bar W$ we have
\begin{align}
    \label{eq:side_bands}
&    \alpha_{xx}(\omega) = \alpha_\mathrm{zr}(\omega)  + 
   \pi \,\mathcal{S}(\Delta\omega)[\bar n(\Delta\omega) +1]
    + \pi\, \mathcal{S}(-\Delta\omega)\bar n(-\Delta\omega), \nonumber\\
& \mathcal{S}(\omega) =    \sum_\qb \frac{|V_\qb|^2}{(\hbar\omega_q)^2} e^{-2\eta_\qb} \,(l_x q_x)^4\delta(\omega-\omega_q), \quad
 \Delta\omega = \omega - \Omega_\mathrm{zr}.
\end{align}
This equation explicitly shows that there are two sidebands, $\mathcal{S}(\pm \Delta\omega)$. The sidebands are continuous spectra with typical width $\omega_{q_m}$. The higher-frequency sideband, $\Delta\omega>0$, comes from absorption where the electron and a ripplon are excited by the radiation, whereas the  lower frequency sideband, $\Delta\omega<0$, correspond to the process  where the electron is excited but a ripplon is absorbed. 

As seen from Eqs.~(\ref{eq:zero_ripplon}) and (\ref{eq:side_bands}), there holds the relation
\[\bar W = \int_0^\infty  d\omega \, \mathcal{S}(\omega)[2\bar n(\omega)+1]. %\quad \bar n(\omega) = [\exp(\hbar\omega/k_BT)-1]^{-1}.
\]
It has a simple meaning. As seen from Eq.~(\ref{eq:absorption_general}), $\int_{-\infty}^{\infty} d\omega\,\alpha_{xx}(\omega) = \pi$ independent of the coupling to ripplons, whereas, for weak coupling $\int d\omega  \,\alpha_\mathrm{zr}(\omega) \approx \pi (1-\bar W)$. The reduction of the absorption in the zero-ripplon line is compensated by the sideband absorption.

%%%%%%%%%%%%%%%%%%%%%%%%%%%%%%%%%%%%

\subsection{Explicit expressions in the case of electrons on helium}
\label{subsec:explicit}

The frequencies of ripplons coupled to the intradot electron are low. Therefore the parameters in the expressions for the spectra should be evaluated assuming that $k_BT\gg \hbar\omega_{q_m}$. In the following, to simplify the estimates we set $l_x = l_y$. Then, using the explicit form of the dispersion law $\omega_q = (\sigma_\mathrm{He}q^3/\rho_\mathrm{He})^{1/2}$, where $\sigma_\mathrm{He}$ is the surface tension, from Eq.~(\ref{eq:V_q}) we obtain for the contributions $\bar W_{E_\perp}$ and $\bar W_\mathrm{pol}$ of the pressing electric field and the polarization coupling to $\bar W$ in the form, respectively,  
\begin{align}
    \label{eq:bar_W_He}
    \bar W_{E_\perp} = \frac{3e^2 E_\perp^2}{32 \sqrt{\pi}\hbar^2}\,\frac{l_x^3\rho_\mathrm{He}}{\sigma_\mathrm{He}^2}k_BT, \qquad \bar W_\mathrm{pol} = \frac{9}{512\sqrt{\pi}}\frac{\Lambda^2 k_B T\rho_\mathrm{He}}{\hbar^2\sigma^2_\mathrm{He}l_x}C_\mathrm{pol},
\end{align}
where
\[C_\mathrm{pol}\approx c_\mathrm{pol}{}^2 + 0.70c_\mathrm{pol} + 0.25,\qquad c_\mathrm{pol} = 1-\ln(4l_x/r_B).  
\]
There is also a contribution to $\bar W$ from the cross-term, which is $\propto E_\perp \Lambda$. We do not give an explicit expression for the corresponding term, it is smaller than the sum $\bar W_{E_\perp} + \bar W_\mathrm{pol}$. As noted earlier, $r_B$ in the expression for $\bar W_\mathrm{pol}$ becomes a variational parameter in the presence of $E_\perp$, it is reduced from its $E_\perp=0$-value.  Numerically, for $T=20$~mK and the transition frequency $\Omega_x/2\pi = 4$~GHz we have $\bar W_\mathrm{pol} \lesssim 0.05$, whereas $\bar W_{E_\perp} \lesssim 0.4$ for $E_\perp = 100$~V/cm. 

The contributions $ \mathcal{S}_{E_\perp}$  and $\mathcal{S}_\mathrm{pol}$ to the sidebands from the pressing field and the polarization coupling to ripplons are, respectively,
\begin{align}
    \label{eq:sideband_He}
 &   \mathcal{S}_{E_\perp}(\omega) = \frac{e^2 E_\perp^2}{16 \pi\hbar}\,\frac{l_x^4\rho_\mathrm{He}^{4/3}}{\sigma_\mathrm{He}^{7/3}}\exp[-l_x^2(\rho_\mathrm{He}\omega^2/\sigma_\mathrm{He})^{2/3}]\omega^{2/3} \nonumber\\
&\mathcal{S}_\mathrm{pol}(\omega)= \frac{1}{64\pi}\frac{\Lambda^2 l_x^4 \rho_\mathrm{He}^{8/3}}{\hbar\sigma_\mathrm{He}^{11/3}}\exp[-l_x^2(\rho_\mathrm{He}\omega^2/\sigma_\mathrm{He})^{2/3}]\,
\tilde C_\mathrm{pol}\, \omega^{10/3} 
\end{align}
where $\tilde C_\mathrm{pol} = [1-\ln(4/q_\omega r_B)]^2$, with $q_\omega = (\rho_\mathrm{He}\omega^2/\sigma_\mathrm{He})^{1/3}$. 

It follows from Eqs.~(\ref{eq:side_bands}) and (\ref{eq:sideband_He}) that the sidebands $\mathcal{S}(\pm\Delta\omega)\bar n(\pm \Delta\omega)$  due to the pressing-field induced and the polarization coupling have qualitatively different shapes. Near the zero-ripplon line, the term $\propto \mathcal{S}_\mathrm{pol}$ increases from zero  as $|\Delta\omega|^{7/3}$ with the increasing distance from the line $\Delta\omega$, for small $|\Delta\omega|$, that is, the zero-ripplon line is well-separated from this sideband. In contrast, the sideband $\propto \mathcal{S}_{E_\perp}$ falls off as $|\Delta\omega|^{-1/3}$, that is, it has the form of a tail of the zero-ripplon line. This ``tail'' is not related to the decay of the electron states but rather to the modulation of the electron transition frequency by ripplons.

%%%%%%%%%%%%%%%%%%%%%%%%%%%%%%%%%%%
\section{Single-qubit gates for electrons on helium}
\label{sec:short_pulses}

An important application of quantized intradot electron states is using them as charge qubit states. This is made possible by the long lifetime of the excited electron states and  by the comparatively strong nonparabolicity of the confining in-plane potential, which can be inferred already from Fig.~\ref{fig1}. The parameters $|V_{ij}|$ of the electron Hamiltonian $H_\mathrm{nonlin}$, Eq.~(\ref{eq:H0}), while small compared to $\Omega_{x,y}$, can be much larger than the electron decay rates and the ripplon-induced fluctuations of the electron energy levels. Therefore, radiation with frequency close to $\Omega_x$ causes interstate transitions $\ket{0_x,0_y} \to \ket{1_x,0_y}$, but does not excite transitions  $\ket{1_x,0_y} \to \ket{2_x,0_y}$. The electron system can be then thought of as a quantum 2-level system, with the states $\ket{0}\equiv \ket{0_x,0_y}$ and $\ket{1}\equiv \ket{1_x,0_y}$.  

Single-qubit gate operations can be performed by short resonant radiation pulses. Coupling to ripplons affects the gate fidelity, the primary effect being the modulation of the transition frequency. We will analyze some consequences of this effect by adding to the Hamiltonian (\ref{eq:full_H}) the term $H_p$ that describes a microwave pulse, which  has an electric field inside the dot ${\bf E}_p$ that is polarized along the $x$-axis. The pulse has a time-dependent amplitude $E_p(t)$, and its frequency $\omega_p$ is close to $\Omega_x$, 
\begin{align}
    \label{eq:H_d}
    H_p = -eE_p(t) x\cos(\omega_p t+\phi_p) \approx 
    -\hbar f_p(t) a_x e^{i\omega_p t} - \hbar f_p^*(t) a_x^\dagger e^{-i\omega_p t},  
\end{align}
Here $f_p(t) = e(l_x/2\hbar)E_p(t)\exp(i\phi_p)$, and  we keep in $H_p$ only resonant terms. 

We will consider the effect of a rectangular pulse of duration $t_p$,
\[ f_p(t) = |f_p|e^{i\phi_p}[\Theta(t)- \Theta(t-t_p) ], \]
where $\Theta(t)$ is the step function, and we will use the two-state approximation. Of interest are pulses that lead to a transition $\ket{0} \to \ket{1}$. Therefore $t_p$ is of the order of the reciprocal Rabi frequency, $t_p\sim |f_p|^{-1}$. 

In this setup, the full density matrix $\hat\rho_\mathrm{er}$ of the electron-ripplon system has four nontrivial matrix elements 
$\bra{m_x,0_y}\hat\rho_\mathrm{er}\ket{n_x,0_y}$
 where  $m_x,n_x$ can take on the values 0 and 1. These matrix elements are operators with respect to ripplons. Of interest is their trace over ripplons, $\tr_r\bra{m_x,0_y}\hat\rho_\mathrm{er}\ket{n_x,0_y}$; the ripplons are assumed to be in thermal equilibrium.

It is convenient to analyze the density matrix using the polaronic transformation
\begin{align}
    \label{eq:polaronic}
    U_P = \exp\left[\sum_\qb\frac{\exp(-\eta_\qb)}{\hbar\omega_q}(1- l_x^2q_x^2\,a_x^\dagger a_x)
    (V_\qb b_\qb - V^*_\qb b^\dagger_\qb)
    \right].
\end{align}
After the transformation, the projection  of the Hamiltonian $H$, Eq.~(\ref{eq:full_H}) (i.e., the Hamiltonian in the absence of the drive), on the states $\ket{m_x,0_y}$ takes the form
\begin{align*}
 %   \label{eq:reduced_H}
U_P^\dagger H U_P \to \hbar\Omega_\mathrm{zr} a_x^\dagger a_x + \frac{1}{2}\tilde V_{xx} a_x^\dagger{}^2 a_x^2 + H_r.
\end{align*}
Here we  used that the relevant vibrational  states of the $x$-mode are $\ket{0}$ and $\ket {1}$; for completeness we included the term $\propto   a_x^\dagger{}^2 a_x^2$, which  is multiplied by the renormalized nonlinearity parameter $\tilde V_{xx}$. We note that, as a result of the transformation, the mode frequency has been changed from $\Omega_x$ to $\Omega_\mathrm{zr}$. 

Further, we go to the rotating frame using the transformation $U_\mathrm{RF} = \exp[-it(\omega_pa_x^\dagger a_x + \hbar^{-1}H_r)]$. As a result, we obtain a system of four operator equations for the elements of the density matrix on the electron states,
\begin{align}
    \label{eq:rho_tilde_general}
\rho_{mn}= \bra{m}U_\mathrm{RF}^\dagger U_P^\dagger \hat \rho_\mathrm{er} U_P U_\mathrm{RF} \ket{n}, \quad m,n\in\{0,1\}. 
\end{align}
They read:
\begin{align}
    \label{eq:rho_eqns_complete}
    &\dot{\rho}_{00} = -i\left(f_p^*e^{\xi^\dagger}\,\rho_{01} -f_p e^\xi\rho_{10}\right),
    \qquad \dot{\rho}_{11} = i\left(f_p^*e^{\xi^\dagger}\,\rho_{01} -f_p e^\xi\rho_{10}\right),\nonumber\\
    &\dot{\rho}_{01} = -i\Delta_p \rho_{01} 
     -if_pe^\xi(\rho_{00} - \rho_{11}),\quad \rho_{10} =(\rho_{01})^\dagger,
     \end{align}
     where
     \begin{align}
         \label{eq:xi_operator}
     \xi\equiv \xi(t) = - \sum_\qb\frac{\exp(-\eta_\qb)}{\hbar\omega_q} l_x^2q_x^2
    \left(V_\qb b_\qb e^{-i\omega_q t} - V^*_\qb b^\dagger_\qb e^{i\omega_q t}\right), \quad \Delta_p = \omega_p-\Omega_\mathrm{zr}.
    \end{align}
We used the fact that
\[ U_P^\dagger H_p U_P = -\hbar f_p(t)a_xe^{i\omega_pt} \exp[\xi(0)] + \mathrm{H.c.}
\]
The parameter $\Delta_p$ in Eq.~(\ref{eq:rho_eqns_complete}) is the detuning of the drive frequency from the frequency of the zero-ripplon line of the electron in the quantum dot. We assume that $|\Delta_p|\ll \omega_p$. The operator $\xi(t)$ is anti-Hermitian, $\xi^\dagger(t) = -\xi(t)$, and one can check that $\braket{\xi(t) \xi^\dagger(t)} = \bar W$.

%%%%%%%%%%%%%%%%%%%%%%%%%%%%%%%%%%%%

\subsection{The effects of quantum and classical ripplon-induced fluctuations}
\label{subsec:ripplonic_noise}

Of interest to us are Rabi oscillations of the electron in the presence of  coupling to ripplons, i.e., the  evolution of the populations of the states $\ket{0}$ and $\ket{1}$. These populations are given by the traces of $\tr_r\rho_{00}$ and $\tr_r\rho_{11}$ over  ripplonic states. From Eq.~(\ref{eq:rho_eqns_complete}) it is seen that  the time evolution of $\rho_{mn}$ is characterized by the frequency $\sim |f_p|$. Therefore ripplons with frequencies $\omega_{q}\gg |f_p|$ are averaged out. Typically, $|f_p| \ll k_BT/\hbar$. Therefore it is a good approximation to assume that  quantum fluctuations of the ripplonic field as a whole are averaged out; here we take into account that the coupling to low-frequency quantum fluctuations of this field is weak, cf. Eq.~(\ref{eq:sideband_He}). Then in Eq.~(\ref{eq:rho_eqns_complete}) we can replace
\begin{align}
\label{eq:f_Debye_Waller}
&f_p(t)\,e^{\xi(t)}\to \widetilde f_p(t),\qquad \widetilde f_p(t) = \bar f_p 
%\,e^{-\bar W_q}\,
\exp[\xi_T(t)], \nonumber\\
&\bar f_p=f_p \,e^{-\bar W_q}, \quad 
\bar W_q = \sum_\qb \frac{|V_\qb|^2}{(\hbar\omega_q)^2} e^{-2\eta_\qb} \,(l_x q_x)^4\, .%,\nonumber\\
%& \tilde f_p(t) = f_p\exp[\xi_T(t)].
\end{align}
Here, $\bar W_q$ is the contribution to the Debye-Waller factor $\bar W$ at $T=0$, cf. Eq.~(\ref{eq:zero_ripplon}).

The term $\xi_T(t)$ accounts for the  coupling to thermal  ripplons. Typically, their  frequencies $\omega_q$ satisfy the condition $\omega_q \ll k_BT/\hbar$. From Eq.~(\ref{eq:xi_operator}), the correlation function of $\xi_T(t)$ is obtained from the correlation function of $\xi(t)$ by keeping the terms $\propto \bar n_q \approx k_BT/\hbar\omega_q$,
\begin{align}
    \label{eq:xi_corelator}
  \Xi(t-t')=  \braket{\xi_T(t) \xi_T^\dagger(t')} = 2 k_BT\sum_\qb \frac{|V_\qb|^2}{(\hbar\omega_q)^3} e^{-2\eta_\qb} \,(l_x q_x)^4\,\cos[\omega_q(t-t')],
\end{align}
whereas the commutator $\braket{[\xi_T(t),\xi^\dagger_T(t')]}$ is smaller than $\Xi(t-t')$ by a factor $\sim \hbar\omega_{q_m}/k_BT$.
Therefore  $\xi_T(t)=-\xi^\dagger_T(t)$ is essentially a classical zero-mean Gaussian noise with the correlator $\Xi(t)$. This noise affects Rabi oscillations. Tracing over ripplons becomes equivalent to statistical averaging over the realizations of the noise $\xi_T(t)$.
%
%\[\tr_r\rho_{mn} \to \braket{\rho_{mn}}\]
%
%where we use $\braket{\cdot}$ to indicate such averaging.

We note that the time-averaging used to obtain $\bar W_q$ is not limited to the coupling to ripplons.  However, the approximation $\omega_q \ll k_BT/\hbar$ is specific for ripplons, given   the low temperatures used in quantum computing systems. We also note  a simple relation between the noise correlator $\Xi(t)$ and the sideband spectral density $\mathcal{S}(\omega)$,
\[ \Xi(t) = \int_0^{\omega_T} d\omega\,\frac{2k_BT}{\hbar\omega} \,\mathcal{S}(\omega) \cos\omega t, 
\]
where the cutoff $\omega_T$ is  $\lesssim k_BT/\hbar$; as explained above, for electrons on helium one can set $\omega_T\to \infty$, given the exponential falloff of $\mathcal{S}(\omega)$ that typically occurs already for $\omega \ll  k_BT/\hbar$.

%%%%%%%%%%%%%%%%%%%%%%%%%%%%%%%%%%%%%

\subsubsection{Nonergodic response}

The problem of Rabi oscillations can be solved in several limiting cases. One of them is the case of comparatively large drive and, respectively, short pulses $t_p$. If the typical ripplon frequency $\omega_{q_m}$ is small compared to $t_p^{-1}$, i.e., the correlation time of the noise $\xi_T$ is large compared to $t_p$, one can replace $\xi_T(t)$ in Eq.~(\ref{eq:f_Debye_Waller}) for $\widetilde f_p(t)$ with  $\xi_T(0)$, so that $\widetilde f_p(t)$ becomes time-independent. Then Eq.~(\ref{eq:rho_eqns_complete}) becomes a standard equation for Rabi oscillations with a time-independent drive.
The noise determines the static random phase of the drive. If the electron is initially in the ground state and the Rabi frequency $R_\mathrm{nonergodic}$ is extracted from the occupation of the excited state in response to a pulse, from Eq.~(\ref{eq:rho_eqns_complete}) we find
\[R_\mathrm{nonergodic} = \left\{\Delta_p^2 + 4|\bar f_p|^2 \right\}^{1/2}.
\]
Unexpectedly, the Rabi frequency measured this way is independent of the noise. 

However, the off-diagonal matrix elements $\rho_{01} = \rho_{10}^*$ depend on a particular realization of the noise. If the noise is so slow that several measurements of $\rho_{01}$ can be repeated for the same value of $\xi_T$ to average over the quantum measurements outcomes, the result will be nonergodic. It will depend on a particular value of $\xi_T$ and will not include averaging over $\xi_T$, cf.~\cite{wudarski2023a}. In other words, the values of $\rho_{01}$ will be different in different series of such measurements. 

On the other hand, if $\omega_{q_m}t_p\ll 1$, but the noise $\xi_T$ varies from measurement to measurement, one can still assume that $\xi_T$ is constant during a pulse, but the measurement outcomes will have to be averaged over the realizations of $\xi_T$. For example, if the electron is in the ground state at $t=0$ and $\Delta_p=0$, we have from Eq.~(\ref{eq:rho_eqns_complete})
\[\rho_{01}(t) = -i\frac{f_p}{2|f_p|} \exp[\xi_T(0)]\sin (2|\bar f_p|t). 
\]
The results of uncorrelated measurements of $\rho_{01}$ given by this expression have to be averaged over $\xi_T(0)$.

%%%%%%%%%%%%%%%%%%%%%%%%%%%%%

\subsubsection{Response to a resonant pulse for strong electron-ripplon coupling}
\label{subsubsec:pulse_strong}

Nonergodic measurements will be hard to do with electrons in a quantum dot on helium.
A less demanding setting is where $\omega_{q_m}$ is comparable or larger than $t_p^{-1}$. In this case the variation of $\xi_T(t)$ during a pulse has to be taken into account.
We will study the response to a resonant pulse, assuming that at $t=0$ the electron is in the ground state. Tracing over ripplons will be replaced by averaging over the classical noise $\xi_T(t)$. From Eq.~(\ref{eq:rho_eqns_complete}) we obtain an equation for the difference of the state populations $\mathcal{N}(t) = \rho_{00}(t) - \rho_{11}(t)$, which reads
\begin{align}
    \label{eq:popul_diff_eqn}
    &\frac{d}{dt}\mathcal{N}(t) = -2|\bar f_p|^2 \int_0^t dt'
    \,\exp\left[-i\Delta_p(t-t') -\xi_T(t) + \xi_T(t')\right]  
    \mathcal{N}(t') + \mathrm{c.c.}\,.
\end{align}
For strong coupling to ripplons, $\bar W\gg 1$, the mean value of the random factor in Eq.~(\ref{eq:popul_diff_eqn}),
\begin{align}
\label{eq:strong_cplng_pulse_factor}
\braket{\exp[-\xi_T(t) + \xi_T(t')]} = \exp\left[-\Xi(0) + \Xi(t-t')
\right]\approx \exp[-\gamma^2(t-t')^2/2],
\end{align}
rapidly falls off  with the increasing $t-t'$. In the above expression we used Eq.~(\ref{eq:strong_coupling}) for the parameter $\gamma$, in which we replaced $2\bar n_q + 1$ with $2k_BT/\hbar\omega_q$. We also used that $\gamma\gg \omega_{q_m}$ for strong coupling, and therefore $\omega_q|t-t'|\ll 1$ for $|t-t'|\lesssim 1/\gamma$.

If $|\bar f_p|\ll \gamma$, it is seen from Eq.~(\ref{eq:popul_diff_eqn}) that $\mathcal{N}(t)$ slowly varies on the time scale $\gamma^{-1}$. Then in this equation one can replace $\mathcal{N}(t')$ in the integrand with $\mathcal{N}(t)$, which gives
\begin{align}
    \label{eq:pop_diff_strong_cpling}
    \braket{\mathcal{N}(t)} = \braket{\mathcal{N}(0)}\exp(-\Gamma_\mathrm{sc}t), 
    \quad \Gamma_\mathrm{sc} = \frac{\sqrt{8\pi}}{\gamma}|\bar f_p|^2\exp(-\Delta_p^2/2\gamma^2).
\end{align}
As seen from this equation, rather than experiencing Rabi oscillations, the difference of the state populations exponentially decays in time, that is, the populations of the state $\ket{0}$ and $\ket{1}$ approach each other.  The decay rate in the strong-coupling limit is $\Gamma_\mathrm{sc} \ll |\bar f_p|$. This rate falls off with the increasing strength of the coupling to ripplons, which determines $\gamma$, and in particular with the increasing temperature. It is quadratic in the amplitude of the radiation pulse $\propto |\bar f_p|$. 

The condition $\Gamma_\mathrm{sc} \ll \gamma$, which underlies replacing $\mathcal{N}(t')$ with $\mathcal{N}(t)$ in Eq.~(\ref{eq:popul_diff_eqn}), is met ``automatically'', since $|\bar f_p|\ll \gamma$. The corresponding inequalities also justify the  next step, in which $\mathcal{N}(t)$ is replaced with  $\braket{\mathcal{N}(t)}$.  

Equation (\ref{eq:pop_diff_strong_cpling}) shows that, practically, one cannot perform coherent gate operations on a charge qubit if the coupling of the underlying electron to ripplons is strong, at least unless the microwave field is so strong that $f_p\gg \gamma$ and the operations are done over time smaller than $\gamma^{-1}$. The constraint is not related to the lifetime of the excited electron state. It also may not be directly mapped on qubit decoherence described by the standard $T_2$-type process.

%%%%%%%%%%%%%%%%%%%%%%%%%%%%%%%%%%%%%%%%%

\subsubsection{Response to a resonant pulse for weak coupling to ripplons}
\label{subsubsec:weak_cpling}

We now consider the effect of a radiation pulse in the case of weak electron-ripplon coupling. It is important to make sure that the divergence of $\mathcal{S}(\omega)/\omega$ for $\omega\to 0$ does not make it impossible to perform reliable single-qubit gates. Moreover, it is necessary to estimate the gate error related to the coupling to ripplons.

We will again assume that initially ($t=0$) the qubit is in the ground state, $\braket{\mathcal{N}(0)} = 1$, and that $\Delta_p=0$, i.e., the pulse frequency $\omega_p$ is equal to the zero-ripplon transition frequency $\Omega_\mathrm{zr}$.  We will find the lowest-order in $\xi_T$ correction to $\mathcal{N}$. To do it, we rewrite  Eq.~\ref{eq:popul_diff_eqn}) in the form
\[\frac{d^2}{dt^2}\mathcal{N} + 4|\bar f_p|^2 \mathcal{N} =
2|\bar f_p|^2\frac{d}{dt}\xi_T(t)\int_0^t dt' \mathcal{N}(t') \exp[-\xi_T(t) + \xi_T(t')] + \mathrm{H.c.}
\]
The zeroth-order solution $\mathcal{N}\0(t)$ of this equation describes Rabi oscillations, 
\begin{align}
\label{eq:N_0}
\mathcal{N}\0(t) = \cos(2|\bar f_p|t).
\end{align}
The correction to $\mathcal{N}(t)$ of the lowest order in $\xi_T$ has the form
\begin{align}
\label{eq:N_1}
&\mathcal{N}\1(t) =\frac{2|\bar f_p| k_B T}{\hbar}\int_0^\infty d\omega \frac{\omega\,\mathcal{S}(\omega)}{(\omega^2 - 4|\bar f_p|^2)^2}\nonumber\\
&\times\left\{[4|\bar f_p|[\cos(\omega t) - \cos(2|\bar f_p|t)] +(\omega^2 - 4|\bar f_p|^2)t\,\sin(2|\bar f_p|t)\right\} 
\end{align}
This equation shows that the correction to the Rabi-oscillation term (\ref{eq:N_0}) increases with the increasing temperature. It is seen that the major contribution to $\mathcal{N}\1(t)$ comes from the range of the frequencies of the sideband spectrum  $\mathcal{S}(\omega)$ where  $\omega\sim 2|\bar f_p|$, i.e., from the range  where the noise ``resonates'' with the Rabi oscillations; the contributions from the low- and high-frequency parts of $\mathcal{S}(\omega)$ are suppressed. Interestingly, the function $\mathcal{N}\1(t)$ displays oscillations at the Rabi frequency with an amplitude that increases with time.   

Overall, Eq.~(\ref{eq:N_1}) shows that the correction to the qubit dynamics is small for weak coupling. However, the form of the correction is somewhat unexpected. The explicit expression for  $\mathcal{N}\1(t)$ makes it possible to find an optimal value of the Rabi frequency $2|\bar f_p|$ for a  desired gate operation.

%%%%%%%%%%%%%%%%%%%%%%%%%%%%%%%%%%%%%%%%%%

%%%%%%%%%%%%%%%%%%%%%%%%%%%%%%%%%%%%%%%%%%
\section{Discussion}
\label{sec:discussion}

The results of this paper show that electrons in quantum dots on helium surface provide an extremely rich system to explore. Their energy spectrum is discrete and can be controlled electrostatically. When the system is embedded into a microwave cavity, one can observe resonant transitions between the energy levels. This underlies using the electron states as charge qubit states.
However, even though the system is free from defects, the electron coupling to the quantum field of  ripplons may pose an obstacle to implementing charge qubits. The physics here is different from the conventional analysis of the processes that lead to electron relaxation. It is rather related to the physics encountered in the studies of color centers in solids. 

The effect of the coupling on  gate operations can be reduced to an effective Debye-Waller factor, which is due to quantum fluctuations and  decreases the amplitude of the driving field, and a classical noise that modulates this amplitude. Such description is possible because, for typical parameters of quantum dots on helium, the frequencies of the  ripplons coupled to an electron are small compared to $k_BT/\hbar$. The slowness of the ripplons may lead to nontrivial measurement outcomes, particularly if the measurements are fast.

Further  insight can be gained using the  explicit expression for the characteristic dimensionless parameter of the electron-ripplon coupling that we provide.  
If this parameter is large, it means that the coupling is strong. In this regime, practically, one cannot perform coherent gate operations on an electron charge qubit. The coupling depends on the electron localization length in the dot, the temperature, and most importantly, the field $E_\perp$ that presses the electrons against the helium surface. This field is necessary to prevent ``evaporation'' of the electrons from the surface. As we show, for the coupling parameter to be small, $E_\perp$ must be  comparatively  small itself, $\lesssim 100$~V/cm. For weak coupling, the error of single-qubit gate operations can be small; it can be estimated using the explicit expressions provided in the paper.

The possibility to vary the coupling to ripplons from weak to strong opens a unique way of studying physics of color centers formed by electronic defects in solids. The electron-ripplon coupling mimics the electron-phonon coupling in solids. It is known that the spectral lines of color centers have a complicated structure, but it has never been possible to study how the lines change as the coupling parameters are varied. For electrons on helium, the variability is enabled by $E_\perp$. Our results show how the electron absorption spectrum changes with decreasing $E_\perp$ from a comparatively broad  Gaussian peak with a superposed weak zero-ripplon line, for large $E_\perp$, to a strong narrow zero-ripplon peak, for small $E_\perp$, with sidebands that have a characteristic form. 

The linear and nonlinear response of a localized electron to a resonant electromagnetic field and the change of this response with the varying control parameters open a way to a better understanding of the electron coupling to a quantized field of helium vibrations, including the features of the coupling in confined geometries. It also provides a means to study the overall structure of the electron motional states in a quantum dot on the helium surface. In a broader context, the system offers a controlled platform for studying the effects of electron coupling to a bosonic field.

%%%%%%%%%%%%%%%%%%%%%%%%%%%%%%%%%%%%%%%%%%

%%%%%%%%%%%%%%%%%%%%%%%%%%%%%%%%%%%%%%%%%%

%%%%%%%%%%%%%%%%%%%%%%%%%%%%%%%%%%%%%%%%%%
\authorcontributions{The  authors have contributed equally and have read and agreed to the published version of the manuscript.}

\funding{This work was supported in part by the National Science Foundation via grant number DMR-2003815.  MID acknowledges partial support from the Google Faculty Award and and from the Gordon and Betty Moore Foundation Award No. GBMF12214. JP acknowledges partial support from  the Cowen Family Endowment at MSU.}

%\acknowledgments{}

\conflictsofinterest{MID declares no competing interests. JP is a co-founder and CSO of EeroQ Corp. }  

%%%%%%%%%%%%%%%%%%%%%%%%%%%%%%%%%%%%%%%%%%
%% Optional

%% Only for journal Encyclopedia
%\entrylink{The Link to this entry published on the encyclopedia platform.}

%%%%%%%%%%%%%%%%%%%%%%%%%%%%%%%%%%%%%%%%%%
%% Optional
%\appendixtitles{no} % Leave argument "no" if all appendix headings stay EMPTY (then no dot is printed after "Appendix A"). If the appendix sections contain a heading then change the argument to "yes".

\reftitle{References}

%\bibliography{MyLibrary,JP1}

% \bibliography{your_external_BibTeX_file}

%=====================================
% References, variant B: internal bibliography
%=====================================

% ACS format

% APA format (Used for journal: admsci, behavsci, businesses, econometrics, economies, education, ejihpe, games, humans, ijfs, journalmedia, jrfm, languages, psycholint, publications, tourismhosp, youth)
\isAPAStyle{%

}{}

% If authors have biography, please use the format below
%\section*{Short Biography of Authors}
%\bio
%{\raisebox{-0.35cm}{\includegraphics[width=3.5cm,height=5.3cm,clip,keepaspectratio]{Definitions/author1.pdf}}}
%{\textbf{Firstname Lastname} Biography of first author}
%
%\bio
%{\raisebox{-0.35cm}{\includegraphics[width=3.5cm,height=5.3cm,clip,keepaspectratio]{Definitions/author2.jpg}}}
%{\textbf{Firstname Lastname} Biography of second author}

% For the MDPI journals use author-date citation, please follow the formatting guidelines on http://www.mdpi.com/authors/references
% To cite two works by the same author: \citeauthor{ref-journal-1a} (\citeyear{ref-journal-1a}, \citeyear{ref-journal-1b}). This produces: Whittaker (1967, 1975)
% To cite two works by the same author with specific pages: \citeauthor{ref-journal-3a} (\citeyear{ref-journal-3a}, p. 328; \citeyear{ref-journal-3b}, p.475). This produces: Wong (1999, p. 328; 2000, p. 475)

%%%%%%%%%%%%%%%%%%%%%%%%%%%%%%%%%%%%%%%%%%
%% for journal Sci
%\reviewreports{\\
%Reviewer 1 comments and authors’ response\\
%Reviewer 2 comments and authors’ response\\
%Reviewer 3 comments and authors’ response
%}
%%%%%%%%%%%%%%%%%%%%%%%%%%%%%%%%%%%%%%%%%%
\PublishersNote{}
%\isPreprints{}{% This command is only used for ``preprints''.
%\end{adjustwidth}
\end{document}